\documentclass[12pt,preprint]{aastex}

\newcommand{\ltsimeq}{\raisebox{-0.6ex}{$\,\stackrel
        {\raisebox{-.2ex}{$\textstyle <$}}{\sim}\,$}}

\revised{}
\accepted{}

\shortauthors{Boyer et al.}
\shorttitle{``Spitzer Detection of the ICM in M15''}

\received{2006 April 24}
\begin{document}

\title{Stellar Populations and Mass-Loss in M15: \\ A 
\textit{Spitzer} Detection of Dust in the Intra-Cluster Medium}

\author{Martha L.\ Boyer\altaffilmark{1,2}, Charles E.\ Woodward\altaffilmark{2}, 
Jacco Th.\ van Loon\altaffilmark{3}, Karl D. Gordon\altaffilmark{4}, A. Evans\altaffilmark{3}, Robert D.\ Gehrz\altaffilmark{2},
L.\ Andrew Helton\altaffilmark{2}, Elisha F.\ Polomski\altaffilmark{2}}

\altaffiltext{1}{mboyer@astro.umn.edu}
\altaffiltext{2}{Astronomy Department, School of Physics and Astronomy, 116
  Church Street, S.E., University of Minnesota, Minneapolis, MN 55455}
\altaffiltext{3}{Astrophysics Group, School of Physical \& Geographical
 Sciences, Keele University, Staffordshire ST5 5BG, UK} 
\altaffiltext{4}{Steward Observatory, University of Arizona, 933 North Cherry
  Avenue, Tucson, AZ 85721}



\begin{abstract}

We present \textit{Spitzer Space Telescope} IRAC and MIPS observations of the 
galactic globular cluster M15 (NGC 7078), one of the most metal-poor 
clusters with a [Fe/H] = -2.4.  Our \textit{Spitzer} images reveal a 
population of dusty red giants near the cluster center, a previously 
detected planetary nebula (PN) designated K648, and a possible detection 
of the intra-cluster medium (ICM) arising from mass loss episodes from
the evolved stellar population.  Our analysis suggests 
$(9 \pm 2) \times 10^{-4}$~M$_{\odot}$ \ of dust is present in the core of 
M15, and this material has accumulated over a period of $\approx 10^{6}$~yrs,
a timescale ten times shorter than the last galactic plane crossing event.
We also present \textit{Spitzer} IRS follow up observations of
K648, including the detection of the [Ne~II] $\lambda 12.81$~\micron \ line,
and discuss abundances derived from infrared fine structure lines.

\end{abstract}


\clearpage
\vfill\eject 
\section{INTRODUCTION}
\label{sec:intro}

Mass loss in evolved stellar populations affects the chemical evolution of 
the interstellar medium (ISM) and mass loss from individual stars governs post 
main sequence evolution. The amount and duration of mass loss that occurs 
in giant stars remains one of the most uncertain parameters in stellar 
evolution theory and the effect of these processes on the inferences 
derived from stellar population models can be significant. Given the wide 
use of these models (e.g., in inferring stellar masses of high-redshift 
galaxies), understanding the mass losing process is vital for a range of 
problems in astrophysics. Although dust constitutes a small fraction of 
the total mass lost, it is frequently used as a marker as it is
optically thin and its thermal 
emission is readily detectable in a variety of environments. 

Globular clusters (GCs), believed to have formed during the assemblage of
the Galaxy, are coeval samples of stars at common, well-determined
distances with nearly uniform initial compositions. GCs enable study
of the chemical enrichment of the interstellar medium arising from mass
ejection during the post-main sequence evolution of stars. Red Giant
stars, especially those ascending the Asymptotic Giant Branch, are
expected to develop winds that inject processed material into the
intra-cluster medium (ICM) during post-main sequence evolution.  These
winds contain gas and solid phase materials, the latter in the form of
dust grains that condense from the metals. However, detection of thermal
emission arising from intra-cluster medium dust has been elusive,
suggesting that the ICM in GCs is 100 to 1000 times less massive than
expected from current stellar evolution theory and observations of
mass-losing stars in clusters and in the solar neighborhood.

The circumstellar environments of stars in the late stages of evolution, 
when most mass loss occurs, are most effectively detected and studied in 
the infrared (IR). IR surveys conducted by {\it IRAS}, {\it 2MASS,} and 
{\it ISO} revealed populations of dust-enshrouded asymptotic giant branch 
(AGB) and red supergiant (RSG) stars in the Galactic bulge 
\citep{Jacco03}, the Large Magellanic Cloud (LMC) 
\citep{Zijlstra96, jacco97, trams99}, and in 
galactic Globular Clusters \citep{Ramdani01,origlia02}. 
GCs are 
expected to contain dust from episodes of mass loss in red giant
branch (RGB) and AGB 
stars. The IR excess of their circumstellar dust emission is expected to 
peak between 20~\micron{} to 30~\micron{}, and thus photometry at 
wavelengths larger than 20~\micron{} is necessary to estimate accurate 
dust masses lost by such stars. The amount of dust present in the 
intra-cluster medium (ICM) will vary depending on the cluster escape 
velocity, the time since last crossing of the Galactic disk where the
ICM can be stripped away by the ISM, and the 
number of mass-losing stars. In general the dust in the 
ICM of GCs is expected to be 
$\simeq 10^{-2}$ to $10^{-3}$ M$_{\odot}$ for most galactic 
clusters. Globular clusters 
have reasonably homogeneous (in age and metallicity), well-understood 
stellar populations, so observations of ICM dust are reasonably 
straightforward to interpret to yield mass loss rates and duty cycles. 

Previous attempts to detect the ICM in GCs suggest that the ICM density is 
well below that expected from predictions of the mass loss input from
RGB and AGB stars, even considering the low metallicity of GCs. The
lowest 3-$\sigma$ upper limits to the ICM mass for 70~K dust are $\sim
6\times10^{-5}$ M$_\odot$ \citep{hopwood99}. Detecting thermal
emission from the elusive ICM in GCs is observationally
challenging. \citet{origlia02} reported \textit{ISO} observations of
the IR thermal emission from the winds of individual RGB stars in six
massive GCs (47 Tuc, NGC 362, Omega Cen, NGC 6388, M15, and M54)
showing that, in those systems, stellar winds from these stars are
enriching the ICM.  Though thermal emission from the ICM material
might be expected to be detectable, many attempts to do so with IR and
millimeter observatories have produced only a single secure detection
of ICM dust, a tentative (3.5-$\sigma$) detection of thermal emission
in the core of the  metal-poor GC M15 \citep{evans03}.  Overall this
result suggests that the ICM dust in GCs is significantly less massive
than expected from current stellar evolution theory and observations
of mass losing stars in GCs and the solar neighborhood. 

The causes of the paucity 
of ICM emission have been proposed to be ram-pressure stripping of ICM gas 
during Galactic plane passage, blowout by nova explosions, fast winds from 
the stars themselves, radiative ejection by the sheer luminosity of 
cluster stars, and continuous ram-pressure from hot gas in the
galactic halo.  However, the dominant state of the ICM is unclear. 
Sensitive searches for neutral H in the ICM at radio wavelengths have yielded upper 
limits  $\leq$ 0.1 M$_\odot$ \citep{Birk83}, with a possible detection of 
$\approx$ 200 M$_\odot$ in NGC 2808 \citep{Faulk91} and a
5$\sigma$ detection of 0.3 M$_\odot$ in M15 \citep{Jacco06}. Perhaps much 
of the ICM is ionized, as suggested by the high electron densities 
measured from pulsar timing in 47 Tuc \citep{Paolo01}. However,
H$\alpha$ searches have as yet been unsuccessful.

ICM dust grains in radiative thermal equilibrium should attain 
temperatures of 50~K to 80~K because of the high energy density of 
starlight within a GC \citep{Forte02} and thus be detectable as an IR 
excess (above the photospheric emission) in GCs at mid- to far-IR 
wavelengths. Here we present observations of the galactic GC M15 with
the \textit{Spitzer Space Telescope} \citep{Werner04}, whose instrumental sensitivity enables
detection of dust masses as low as $4\times10^{-9}$ M$_\odot$
(assuming a population of grains radiating in thermal equilibrium,
with integration times down to the background confusion limit) and
therefore permits an unequalled opportunity to search for and set
stringent limits on the ``missing'' ICM. 

\subsection{M15}
\label{sec:m15}

M15 (NGC 7078), with [Fe/H] = -2.4 \citep{sneden97}, is one of the
most metal poor GCs. It is a well-studied cluster, as it is home to
the first planetary nebula (PN) discovered in a GC (K648, also designated as
Ps-1, \citet{pease28,howard97,alves00}) and to the first GC Low-Mass X-ray Binary source
  (X2127+119, \citet{auriere84,charles86}).  At least
eight millisecond pulsars are also associated with the cluster
\citep{Kulkarni96}.  M15 is generally
believed to be a core-collapse GC, with a small, dense core
containing approximately 4000~M$_\odot$ \citep{phinney96}.

Properties of M15, reproduced from \citet{hopwood99}, are listed in
  Table~\ref{tab:m15_params}. Updated values as listed by
  \citet{evans03} include the escape velocity $V_{esc}$ from
  \citet{webbink85}, and the time $\tau_c$ since the last plane crossing
  from \citet{oden97}.  The reddening and the distance are updated
  from \citet{schlegel98} and \citet{mcnamara04}, respectively.
  Galactic coordinates for M15 are $l = 65.01^\circ$ and $b =
  -27.31^\circ$, placing the cluster $\sim$ 4.5~kpc south of the galactic plane. 

The total dust mass expected in a GC can be estimated using the
following equation:

\begin{equation}
M_{dust} = \frac{\tau_c}{\tau_{HB}} \ N_{HB} \ \delta M \ \frac{10^{[Fe/H]}}{100},
\end{equation}

\noindent where $\tau_{HB}$ is the Horizontal Branch (HB) lifetime, $N_{HB}$ is
the number of HB stars, [Fe/H] is the cluster metallicity, and $\delta
M$ is the dust mass lost from each star at the tip of the RGB. The
factor of 100 is the solar gas-to-dust ratio, which is scaled for the
metallicity of M15 by adding the [Fe/H] factor. Using values typical
of population II stars and this relationship, the expected dust mass
in M15 has been estimated to be 3.7 $\times$ 10$^{-3}$ M$_{\odot}$ by
\citet{evans03} and 2.0 $\times$ 10$^{-3}$ M$_{\odot}$ by
\citet{hopwood99}. 

M15 is also home to the PN K648.  K648 was the 
first globular cluster PN discovered, and subsequently it has been 
extensively studied at UV, optical and IR wavelengths to determine 
the chemical composition of the ejecta nebula and parameters of the central 
star \citep{barker83, adams84, howard97, bianchi01, hld03,
  garnett93}. The study of post-AGB stellar evolution in old metal-poor, low-mass stars 
in GCs or the galactic halo can be greatly enhanced if the 
by-products of stellar nucleosynthesis can be measured. Enriched material 
produced in the RGB and AGB stages of stellar evolution are dispersed 
into outer layers of the stellar system and subsequent mass loss 
processes lead to the formation of a PN. Few PNe in the galactic halo
population have been identified and only four of these PNe, including
K648, BB-1, DdDM~1, and H4-1 are associated with globular clusters
\citep{jacoby97}.  Below, we present findings derived from a 5-15
\micron{}~ IR spectrum of K648.

\section{OBSERVATIONS}
\label{sec:obsec}

Image observations of the GC M15 were obtained on 2004 October 
29 UT with the Multiband Imaging Photometer for \textit{Spitzer} 
(MIPS) \citep{Rieke04} camera through the 24~\micron{} and 70~\micron{} 
filters and with the Infrared Array Camera (IRAC) \citep{Fazio04} at 3.6, 
4.5, 5.8, and 8~\micron{} conducted as part of the Gehrz Guaranteed 
Time Observing Program (GGTOP, PID 124). Raw data were processed with the 
\textit{Spitzer} Science Center (SSC) pipeline version S11.4.0. To avoid 
saturation in IRAC, High Dynamic Range (HDR) mode was implemented, using 
0.4~s, 10.4~s, and 98.6~s exposures.  For each IRAC channel, 55 frames at 
seven dither positions were obtained with each HDR exposure time. With 
the MIPS camera, 596 dithered frames were taken at 24~\micron{} along 
with 256 dithered frames at 70~\micron{}. 

 \textit{Spitzer} IRS spectra 
of the PN and other red sources detected in the IRAC images have been 
obtained as part of a follow on program using GGTOP time and these data are discussed 
below. Table~\ref{tab:obsum} 
summarizes the observations discussed in this paper. 

\subsection{IRAC}

The IRAC Basic Calibrated Data (BCD) images were post-processed to 
correct for various instrumental artifacts and were mosaicked using the 
2005 May 09 version of the SSC Legacy MOPEX software \citep{mopex}.  The MOPEX cosmetic 
correction was used to eliminate mux-bleed and column pull-down as 
described by the IRAC Data Handbook,
v3.0.\footnote{http://ssc.spitzer.caltech.edu/irac/dh/iracdatahandbook3.0.pdf}
The background matching correction was used to minimize pixel
differences in overlapping areas of the mosaics. The MOPEX mosaicker
also eliminated cosmic rays and other outliers in the data. The final
IRAC mosaics are not subsampled, thus the pixel size of each final
tiled image is 1.22\arcsec \ per pixel covering an area $\sim 50$
square arcminutes around the core (RA[J2000] =
21$^{\mbox{\small{h}}}$:29$^{\mbox{\small{m}}}$:58\arcsec.38;
Dec[J2000] = +12$^{\circ}$:10\arcmin :00\arcsec.6) of M15 plus an
off-field region of the same size 7\arcmin{} to the north or south. A
three color image combining 3.6~\micron{}, 4.5~\micron{} and
8~\micron{} is presented in  Fig.~\ref{fig:3color}a. The diffuse ICM
is not detected in the IRAC mosaics, even at 8~\micron{}. However, the
planetary nebula (K648) and several other dusty stellar objects become
quite prominent at 8~\micron. 

Point source photometry was conducted using several stellar 
extraction routines including 
GlimpsePhot\footnote{http://www.astro.wisc.edu/sirtf/}, DAOphot 
\citep{Stets87}, and the astronomical point source extraction (APEX) tool 
contained in the SSC MOPEX package. Severe stellar crowding towards the 
cluster core in the IRAC bands made reliable 
point-response-function (PRF) photometry challenging. The best 
photometric results (95\% agreement to 2MASS K-band fluxes) were
obtained by performing PRF fitting on the BCD images with APEX. Array
location dependent photometric correction weights were applied before
PRF fitting to minimize systematic error in the point source
extraction flux. The images were then corrected with the background
matching routine, and outliers were eliminated by the mosaicker. 

PRFs were created for 3.6~\micron{} and 
4.5~\micron{} data using the PRF estimate routine provided in APEX.  At 
least twenty stars were chosen to make each PRF in areas with the 
smallest amount of crowding.  The PRFs provided by the \textit{Spitzer} 
Science Center
(SSC)\footnote{http://ssc.spitzer.caltech.edu/irac/psf.html} that were
made in-flight in January 2005 for 5.8~\micron{} and 8~\micron{} are
sufficiently over-sampled, so there was no need to create new PRFs for
these channels.  Probable point sources with fluxes at least
4-$\sigma$ above the background were then identified and selected for
extraction on coadded images that had been corrected for array
distortion. Final flux extractions were performed at these world
coordinates positions on the corrected BCD images using PRF
subtraction.

After applying the appropriate photometric color correction (as
discussed in the IRAC Data Manual, v3.0) and comparing APEX
3.6~\micron{} fluxes with 2MASS K-band fluxes \citep{2MASS}, the
average [K - 3.6~\micron{}] color is that of a K0 giant star.  If we then
assume that the majority of the stars we detect with {\it Spitzer} are K0
giant stars, then our fluxes are consistent with 2MASS K-band fluxes to
within approximately 5\%.  To further check our photometric results,
we compared the fluxes from thirty of the stars we detect to K-band fluxes from
\citet{cohen05}.  The median [K-3.6~\micron{}] color for these sources
is 0.09, which is also consistent with K0 giant stars (See Fig.~\ref{fig:cohen}).

Color-magnitude diagrams (CMDs) derived from the IRAC point 
source photometry are presented in Fig.~\ref{fig:irac}. 
An RGB is clearly evident with the tip occurring near F$_{3.6}$ = 14
mJy. Stars with 3.6~\micron{} flux $>$ 11.2~mJy are possibly
saturated in the long exposure frames, so only medium exposure (10.4
second) BCDs were used for the PRF subtractions and flux
extraction. Flux densities for stars with colors redward of the RGB
are listed in Table \ref{tab:irsources} (See \S\ref{sec:stars}).

\subsection{MIPS}

The MIPS Data Analysis Tool (DAT, \citet{Gordon05}) version 2.96 was
used to do the basic processing and final mosaicking of 
the individual MIPS images.  In addition, extra processing steps on each 
image were carried out before mosaicking using programs written 
specifically to improve the removal of MIPS detector instrumental 
signatures.  At 24~\micron{} the extra steps included readout offset 
correction, scan mirror dependent flat fields, a scan mirror independent 
flat field, array averaged background subtraction, and exclusion of the 
bias boost images.  At 70~\micron{} the extra steps were column average 
subtraction and pixel time filtering both with the exclusion of the 
regions around the bright sources. The pixel sizes of the final mosaics 
are 1.245\arcsec \ and 4.925\arcsec \ for 24~\micron{} and 70~\micron{}, 
respectively. The 24~\micron{} mosaic covers a 90 square arcminute area 
and the 70~\micron{} mosaic covers a 5.0 arcminute $\times$ 10.0 
arcminute area, each centered on the core of M15. 

The MIPS mosaics (Fig.~\ref{fig:3color}b,\ c) show a possible ICM detection at 
24~\micron{} and 70~\micron. The 24~\micron{} image shows two high 
surface brightness patches of material, both offset from the core by 
$\simeq 17$\arcsec \ towards the west (IR1b and IR2).  The 70~\micron{} image shows only 
one high surface brightness patch that is likely an unresolved image of 
both 24~\micron{} regions. The integrated fluxes of the ICM detections at 
both wavelengths were determined using basic aperture photometry in IDL 
v6.0. A 130.5\arcsec \ square aperture centered at RA[J2000] =
21$^{\mbox{\small{h}}}$:29$^{\mbox{\small{m}}}$:56\arcsec.50;
Dec[J2000] = +12$^\circ$:09\arcmin:53\arcsec.32 yields fluxes of
159.4$\pm$0.1~mJy at 24~\micron{} and 691.2$\pm$61.0~mJy at
70~\micron{}.  The 70~\micron{} flux agrees well with that found by
\citet{evans03} with \textit{ISO} observations 
using the same aperture size and position (Table~\ref{tab:obs_sf}). Also 
visible in the MIPS mosaics are the planetary nebula at 24~\micron{} and 
several dust enshrouded stars, some of which are also visible in the
8~\micron{} image, that may or may not be associated with the 
cluster.

Point source photometry was performed on the 24~\micron{} mosaic using StarFinder \citep{diolaiti00},
which is well suited for the stable and well sampled MIPS 24~\micron{}
PSF.  A STinyTim
\citep{krist02}\footnote{http://ssc.spitzer.caltech.edu/archanaly/contributed/stinytim.tar.gz}
model PSF with a temperature of 100~K and smoothed to account for
pixel sampling was used for the stellar extractions.  It has been
shown that the smoothed STinyTim PSFs are excellent matches to observed MIPS
24~\micron{} PSFs \citep{engelbracht06}.  CMDs comparing 24~\micron{}
to 3.6~\micron{} and 8~\micron{} are presented in Fig.~\ref{fig:mips},
and the uncertainties in the IRAC and MIPS photometry are summarized
in Fig.~\ref{fig:err}.

\subsection{IRS}


Associated with our imaging programs, observations of the PN K648
\citep{pease28,howard97,alves00} in M15 were obtained 
with the \textit{Spitzer} Infrared Spectrograph (IRS) on 2005 November 
17.77~UT using the short wavelength (5-15 $\mu$m), low resolution module 
(SL) in staring mode.  All observations utilized the IRS blue peak up array at 
the target position of the PN, RA[J2000] =
21$^{\mbox{\small{h}}}$:29$^{\mbox{\small{m}}}$:59\arcsec.41;
Dec[J2000] = $+12^{\circ}$:10\arcmin :25\arcsec.70, and the entire H$\alpha$ 
nebulosity of K648 (cf. Fig.~2 of  \citet{alves00}) was contained within 
the spectrograph slit. The slit dimensions are 
57\arcsec~$\times$ 3.6\arcsec, and the slit was oriented 18.92 degrees
west of north. The SL spectroscopic astronomical observing templates
consisted of 5 cycles of 60 second ramps. IRS BCDs were processed with
version 13.0.1 of the IRS pipeline.  A description of the IRS
instrument and its operation is available in \citet{houck04}.  Details
of the calibration and raw data processing are specified in the IRS
Pipeline Description Document,
v1.0.\footnote{http://ssc.spitzer.caltech.edu/irs/dh/PDD.pdf}

Post-pipeline processing was conducted to remove instrumental artifacts, 
perform background subtractions and to combine extracted spectral 
segments. Fatally bad pixels were interpolated over in individual BCDs 
using bad pixel masks provided by the SSC.  Multiple data collection
events were obtained at two different positions on the slit using
\textit{Spitzer's} nod functionality.  The two-dimensional BCDs were
differenced to remove the background flux contribution and then the
data were extracted with the \textit{Spitzer} IRS Custom Extractor
(SPICE)
(v1.1-beta15)\footnote{http://ssc.spitzer.caltech.edu/postbcd/doc/spice.gui\_manual.html}
using the default point source extraction widths.  The extracted,
background corrected data were combined using a weighted linear mean
into a single output data file and clipped at the 3$\sigma$ level.  At
the time of reduction, the errors generated by the SSC pipeline were
not reliable enough for sound interpretation and so the errors were
estimated from the standard deviation of the flux at each wavelength
bin. Where there were less than three points in a wavelength bin, the
error is the quadrature sum of the errors in the files. The resultant
spectrum is presented in Fig.~\ref{fig:pn} and derived line fluxes are
summarized in Table~\ref{tab:k648lf}. The spectral lines were fitted
using a least squares Gaussian routine that fits the line center, line
amplitude, continuum amplitude and the slope of the continuum. The
full-width half-maximum was fixed at the resolution limit of the low resolution
module. Integrating the flux over the 8~\micron{} IRAC bandpass
yields a flux that agrees with the IRAC flux within the uncertainty limits.  The strongest line in the 
mid-IR spectrum is the [\ion{Ne}{2}]$\lambda = 12.81$~\micron{} line, followed 
by hydrogen recombination lines \ion{H}{1}~$6-5$~(Pf$\alpha$)$\lambda = 
7.46$~\micron{} and \ion{H}{1}~$7-6$~(Hu$\alpha$)$\lambda = 12.37$~\micron. 
Emission from S$^{3+}$ is evident in the spectrum, although the fit to
the line 
flux of the [\ion{S}{4}]$\lambda = 10.52$~\micron{} is of marginal 
signal-to-noise (S/N $\approx 2.3$), while no [\ion{Ar}{3}]$\lambda = 
8.99$~\micron \ is seen. Our detections of the mid-IR neon and sulfur 
lines are the first reported in the literature for K648. Abundance 
estimates derived from these forbidden lines are discussed in 
\S\ref{sec:pnk}.

\section{DISCUSSION}
\label{sec:disc}

Our \textit{Spitzer} images of M15 for the first time clearly show both 
the stellar dust producers and the ICM dust (Fig.~\ref{fig:3color}), allowing a direct comparison 
to be made between the dust injection and dust survival rates. The brightest source of 
70~\micron{}  emission, IR1a, is blended at 70~\micron{} but visible as 
separate objects in the MIPS 24~\micron{} map, IR1b and IR2. These sources are 
completely invisible on the IRAC maps, even at
8~\micron{}. A three color image of 8, 24, and 70~\micron{} in which
the 8 and 24~\micron{} images are degraded to match the 70~\micron{}
resolution is
presented in Fig.~\ref{fig:convolved}.  This figure illustrates that IR1a
is not unresolved stellar emission, but is a starless dust cloud(s)
that are likely to be of an intra-cluster nature. The next
brightest objects at 70~\micron{} are a pair of sources, IR3a 
and IR3b, that are situated at the fringes of the cluster and may not be 
physically associated with the cluster. IR3b was previously detected
by 2MASS, but there are no known previous detections of IR3a.  The probability
of detecting non-member red sources in the field are sufficiently
small to suggest that these sources are associated with the cluster.
The only other 70~\micron{} source, IR4, is also located on the fringes of the cluster.
Radial velocity measurements from \citet{pila00} confirm its
membership.  Table~\ref{tab:irsources} lists fluxes for  IR3a, IR3b,
IR4, and other possibly dusty sources, described in
\S\ref{sec:stars}.  All three of these sources have mid-IR colors that
are consistent with post-AGB stars (Fig.~\ref{fig:mips}; \cite{Groenewegen06}). In 
addition to these sources, the planetary nebula K648 (Ps~1) is also detected in all IRAC 
bands and at 24~\micron{}.  The strong 24~\micron{} dectection is
likely due not only to dust, but also to line emission from [\ion{O}{4}] at
25.88~\micron{} and/or [\ion{Fe}{2}] at 25.98~\micron{}.

\subsection{Mass-Losing Stars}
\label{sec:stars}

Figure~\ref{fig:massloss} shows the locations of the mass-losing AGB stars in 
M15. The coordinates and IRAC fluxes of these sources are listed in
Table~\ref{tab:irsources}. These stars were identified by their
locations on the IRAC CMDs 
(Fig.~\ref{fig:irac}), redward of the RGB.  We find 24 
mass-losing/dust-enshrouded stars and consider this a lower limit (due to 
potential source confusion) of the total number in the cluster. These
stars also fall just redward of the RGB in the MIPS CMDs as well,
which suggests that they could be post-AGB stars. Their
IRAC colors, however, indicate that it is
more likely that they are AGB stars that are approaching the end of
the AGB phase of their evolution \citep{Groenewegen06}.  The
stars blueward of the RGB can be explained by the absorption in the
fundamental bands of CO in the 4.5~\micron{} band and SiO in the
8~\micron{} band.  The red, 
mass-losing stars in M15 populate an uneven spatial distribution about the core 
of the cluster as projected on the sky (Fig.~\ref{fig:massloss}). These stars are offset from
the core in the same sense as IR1a,
IR1b, and IR2, and the
distribution is less cusped than the visual light. Since M15 is 13.2
Gyr old \citep{mcnamara04}, most stars currently on the
AGB have a zero-age MS mass of $\simeq 0.8$~M$_\odot$. These stars
will soon end their lives as white dwarfs of $\simeq 0.5$~M$_\odot$, having lost approximately
0.3~M$_\odot$ during their post-main sequence evolution.  The loose 
spatial distribution of this population could therefore be due
to mass segregation, in which lower-mass
stars are displaced to the outer regions
of the cluster due to their high velocities, leaving a preferential concentration of high-mass
stars near the center of the potential well of the cluster \citep{Spitzer87}. 

The MIPS CMDs (Fig.~\ref{fig:mips}) identify a population of 
approximately 23 post-AGB stars \citep{Groenewegen06}. The colors 
of IR3a, IR3b, and IR4 are similar to the
colors of this population. These stars are bright at 24~\micron{}, but
they are not detected above the background at 70~\micron{} 
(Table~\ref{tab:irsources}).  They are distributed around the
cluster center at an average radius of approximately 3.3\arcmin, and their
positions are biassed towards the southern side of the 
cluster, near IR3a, IR3b, and IR4. Mass
segregation is most likely responsible for their locations at the
fringes of the cluster.  

Although 
M15 is one of the most massive galactic globular clusters, it is 
surprising to find many dusty objects, as the cluster metallicity of 
[Fe/H] $= -2.4$ places it amongst the most metal-poor galactic globular 
clusters. Dust forms from metal condensates and it is difficult to 
understand how dust grains can form at such low metal abundances. It
is likely that the metals condensing to form dust are produced in the stars themselves 
and brought to the surface near the end of their evolution. The fact that 
dust production does not seem to be inhibited at metallicities \ltsimeq 
1\% solar implies that stellar mass loss must already have contributed 
dust very early on in the evolution of the Universe. Dust observed at 
high redshift is usually believed to originate in supernovae explosions 
that result from core collapse in massive stars, but our observations 
suggest that at least some of the dust formed within the first few 100 
million years may have been produced by stars of only a few solar
masses (e.g. MS lifetime from models by \citet{vassiliadis93}). 

Various investigators \citep{daulton02, amari01} suggest that
dust grains can form more easily at low metallicity in carbon stars,
as these stars produce carbon themselves. With less initial oxygen
abundance to start with, it is easier for these stars to achieve
C/O~\textgreater~1 in their carbon shells.  With excess carbon available
(after locking up equal amounts of C and O in CO), formation of
carbon-chain molecules from which dust grain condensation can proceed.
However, although there is evidence for high molecular abundances in
stellar atmospheres at
low metallicity, dust production in these stars may
be less efficient due to a lack of SiC seeds \citep{sloan-aph}.
Furthermore, \citet{jacco-aph} suggests that the low optical depth in carbon stars in the
Magellanic clouds points against large dust-to-gas ratios at
low metallicity. 

For low-mass oxygen-rich stars, 
one would not expect this evolution to
lead to effective dust production, as dredge-up only increases the C/O ratio and does not
facilitate the formation of oxygen-rich dust grains.  In any case, for
oxygen-rich stars it is well established that nucleation sites must be
available to condense dust grains onto \citep{jeong99}.  These seeds are likely to be
TiO or similar which include secondary elements that are not produced
by the star itself.  Other s-process seeds, such as Zr, may be
dredged-up in the atmospheres of these stars, but the limiting factor
for dust production with such seeds is the oxygen abundance, as more
oxygen will be locked into CO after dredge-up.  Unfortunately, our
observations do not allow us to draw any conclusions as to the
abundances of secondary elements in the mass-losing stars in M15.

The suspicion is that stars do adjust their structure until they
finally can shed their mantles as demonstrated by K648 in M15. From
the analysis of spectra of GC giant stars, a picture
is emerging in which metal-poor stars do become very cool while
nonetheless exhibiting early-type spectra because the low metal
abundances give rise to weak absorption.  They may not form much dust,
but they may still be able to form enough of it to drive a wind.  The
mid-IR spectrum of 47~Tucanae~V1 suggests typical
silicate dust grains and a typical mass loss rate of
10$^{-6}$~M$_\odot$~yr$^{-1}$ \citep{Jacco06}.

\subsection{IR Spectrum of K648}
\label{sec:pnk}

Although UV and optical line ratios derived for K648 from previous 
investigations provide constraints on relative abundance ratios, several 
important $\alpha$-capture elements, such as S, Ar, and Ne, have ground 
configurations that produce only IR fine-structure. If lines from these ions 
are not observed and introduced into abundance models, the total 
abundance of these elements becomes uncertain \citep{hld03}. Often, 
measuring emission line flux from [\ion{Ne}{2}]$\lambda$12.81~\micron, 
[\ion{S}{4}]$\lambda$10.51~\micron, and [\ion{Ar}{3}]$\lambda$8.99~\micron \ is 
observationally challenging from the ground. However, the sensitivity of 
the \textit{Spitzer} IRS affords an opportunity to set stringent limits 
on the emission line flux of these ions resulting in better constrained 
estimates on derived abundances. In addition, new radiative transition
rates and collision strengths are now available as a 
result of the IRON project \citep{iron_1} to improve the accuracy of derived abundance 
ratios. Below, we discuss the analysis of our \textit{Spitzer} 
measurements (Table \ref{tab:k648lf}) of the [\ion{S}{4}] and [\ion{Ne}{2}] lines in K648 
and present a reanalysis of the S/O and Ne/O ratios with contemporary 
atomic parameters using a simple model. Undertaking a full 
photoionization analysis of K648 (i.e., \citet{howard97}) using the 
\textit{Spitzer} IR line fluxes is beyond the scope of this paper. 

\subsubsection{H$\beta$}
\label{sec:hbeta}

Accurate extinction correction of UV and optical lines with respect to 
H$\beta$ are required for abundance determinations. Interstellar 
extinction estimates to K648 range from $0.2 \leq A_{V} \leq 0.6$ 
\citep{garnett93}. However, at mid-IR wavelengths, extinction is minimal 
($\ltsimeq 0.02$~mag), and the Pf$\alpha$ and 
Hu$\alpha$ lines detections (Fig.~\ref{fig:pn}) enable us to estimate the emitted 
H$\beta$ flux directly. Assuming Case B \citep{osterbrock99}, and 
adopting a $T_{e} = 12,500$~K and density of $\approx 10^{3}~\rm{cm}^{-1}$ 
\citep{garnett93} we derived $F(\rm{H}\beta) = (1.91 \pm 0.30) \times 
10^{-12}$~ergs~cm$^{2}$~s$^{-1}$ and $F(\rm{H}\beta) = (1.13 \pm 0.36) 
\times 10^{-12}$~ergs~cm$^{2}$~s$^{-1}$ from the Pf$\alpha$ and Hu$\alpha$ 
lines respectively using the intrinsic hydrogen emissivity ratios 
tabulated in \citet{humstor87}. These two estimates are in reasonable 
agreement (within the formal error) with each other, considering there is 
an absolute photometric uncertainty of $\approx 5\%$ \citep{houck04} 
between the two spectral orders ($5.2 - 8.7$~\micron{} and $7.4 - 
14.0$~\micron{}). Our average $F(\rm{H}\beta)$ of $(1.52 \pm 
0.23) \times 10^{-12}$~ergs~cm$^{2}$~s$^{-1}$ is 
in good agreement with previous 
observational estimates, especially those obtained with large apertures 
\citep{garnett93}, and we will adopt this value in our abundance 
analysis. 

\subsubsection{Neon}
\label{sec:neona}

The ratio of [\ion{Ne}{2}]$\lambda = 12.81$~\micron{} line flux to our derived 
average value of H$\beta$ was used to estimate the Ne$^{+}$/H$^{+}$ 
abundance. We have adopted a collisional strength $\Upsilon$(T) = 0.283 
(appropriate for T$_{e} = 10^{4}$~K, \citet{sartul94}) and an $A_{if}$ 
value of $8.59 \times 10^{-3}$~s$^{-1}$ from the NIST database and assumed an 
electron density of $N_{e} = 1.7 \times 10^{3}$~cm$^{-3}$ \citep{garnett93}. 
Rate coefficients, $q_{fi}$ \citep{iron_1} and the population levels 
were computed assuming Ne$^{+}$, a $2p^{5}$ ion, is a two-level atom. 
Following \citet{rank78}, we define the relative abundance of 
Ne$^{+}$/H$^{+}$ as 

\begin{equation}
\frac{Ne^{+}}{H^{+}} = \frac{(4\pi\ j_{H\beta}/N_{e}N_{p})N_{e}}
     {h\nu_{fi}A_{if}f_{f}} \times \frac{I([\mbox{\ion{Ne}{2}}])}{I(H\beta)},
\end{equation}

\noindent where the H$\beta$ volume emissivity, $4\pi j_{H\beta}/N_eN_p$, 
is $1.0301 \times 10^{-25}$~erg~cm$^{3}$~s$^{-1}$ interpolated for the 
assumed N$_{e}$ \citep{humstor87}, $f_{f} = 9.86 \times 10^{-4}$ is the 
population ratio of the upper to lower state, and I([\ion{Ne}{2}]), the observed 
neon line flux, is $(1.74 \pm 0.08) \times 10^{-13}$~erg~cm$^{-2}$~s$^{-1}$ 
\ (Table~\ref{tab:k648lf}). This yields a ratio of Ne$^{+}$/H$^{+} = (1.53 \pm 
0.21) \times 10^{-5}$. The relative abundance of Ne$^{2+}$/H$^{+}$ was 
estimated from optical measurements of the [\ion{Ne}{3}]$\lambda 3869$~\AA \ 
and [\ion{Ne}{3}]$\lambda 3967$~\AA \ lines \citep{adams84} and de-reddened 
assuming $c=0.12$ and a \citet{seaton79} extinction law. Using the 
[\ion{Ne}{3}] to H$\beta$ ratios (see Table~\ref{tab:k648abund}), the relative 
populations were computed using a multi-level atom program incorporating 
the best available $A_{ij}$(s$^{-1}$) and collision strengths available 
from the NIST database and the literature \citep{mohesduf04}, which is
similar to code originally described by \citet{shawduf95}. Summing over all ions, we find a 
total Ne/H$ = (2.39 \pm 0.27) \times 10^{-5}$ and a Ne/O $= 0.48 \pm 
0.14$ adopting an O/H $= (5.0 \pm 1.3) \times 10^{-5}$ \citep{pena93}. 

\subsubsection{Sulfur}
\label{sec:sulfa}

The relative sulfur abundance was computed in a similar manner as the 
neon abundance described in \S\ref{sec:neona}, although all relative 
populations were determined using the \citep{mohesduf04} code. The 
\textit{Spitzer} observation of [\ion{S}{4}] was used to determine the 
S$^{3+}$/H$^{+}$ population, while the optical fluxes reported by 
\citep{barker83} corresponding to H$\beta$ (assuming 
$j(H\alpha)/j(H\beta) = 2.81$; \citet{osterbrock99}) were used to 
estimate the relative S$^{2+}$/H$^{+}$ and S$^{+}$/H$^{+}$ abundances 
(Table~\ref{tab:k648abund}). We find an upper limit to the total S/H of 
$(4.28 \pm 1.28) \times 10^{-8}$, which is $\approx 2.5$ lower than that 
inferred by \citet{garnett93}. Adopting an O/H $= (5.0 \pm 1.3) \times 
10^{-5}$ \citep{pena93}, yields a S/O $= (8.56 \pm 3.39) \time 10^{-4}$. 

\subsubsection{Abundance Comments}
\label{sec:nesc}

Our new estimates of the [S/O]$\le -2.64$ and [Ne/O] = +0.54 confirm that 
K648 is under-enriched in S as compared to O \citep{garnett93}, while 
Ne/O is enhanced with respect to solar. The Ne enhancement is seen in 
other halo population PNe, such as BB-1 \citep{hld03}. \citet{garnett93} 
argue that contamination of He-burning products by $\alpha$-captures at 
high temperature could account for enhanced Ne. \citet{bianchi01}, based 
on analysis of \textit{Hubble Space Telescope} (HST)
\textit{Faint Object Spectrograph} (FOS) spectra of the 
central star, suggest that 
the nebular shell was ejected by a low-mass He-burning progenitor that 
has subsequently undergone a late thermal pulse, perhaps similar to the 
evolution of objects akin to FG Sge \citep{gehrz05}. Our derived neon 
abundance using the fine structure line, suggests that dredge-up from 
the stellar core may be an important mechanism to pollute the expelled 
nebular materials from slowly evolving, young PNe.

\subsection{Dust in the ICM}
\label{sec:dust}

Assuming that the diffuse emission detected in the \textit{Spitzer} images 
arises from ICM material, we can compute a dust mass using the observed 
SEDs. The approximate temperature of the ICM dust was derived by 
least-squares fitting a greybody model to the SED using fluxes measured in 
a 130.5 $\times$ 130.5 square arcsecond aperture centered at the same right 
ascension and declination coordinates in images at all wave bands. Our 
choice of aperture size is equivalent to that used by \citet{evans03} to 
facilitate direct comparison.  The large wavelength range of the 
\textit{Spitzer} SED enables us to distinguish between the contribution of 
a stellar blackbody (peaking near 0.64~\micron{} and dominated by K0 
stars), and that of thermal radiating dust that generates an IR excess at 
wavelengths greater than 24~\micron{}. Use of a two-component model, 
incorporating a stellar blackbody that peaks near $4699 \pm 58$~K and a dust 
blackbody that peaks near $T_{d} = 70 \pm 2$~K, gives a rough fit to the data 
(Fig.~\ref{fig:sed}; Table~\ref{tab:obs_sf}). The fit produces a large
reduced $\chi^2$ value of 5.26 suggesting that the integrated
flux within our aperture sums the emission from stars of
many disperate spectral types not well-represented by a simple,
single emissivity and temperature blackbody. Another source of uncertainty in
the fit is the effect of crowding in 2MASS data.  This could lead to
oversubtraction of the background and overestimation of the flux
densities \citep{jacco05}. 

The mass of the ICM was determined using the methodology described in 
\citet{evans03}.  We assume that the dust is optically thin, which yields 
the following expression 

\begin{equation}
\frac{M_d}{M_\odot} = 4.79\times 10^{-17}\ f_\nu (mJy)\  
\frac{D^{2}_{kpc}}{\kappa_\nu B(\nu,T_d)},
\end{equation}

\noindent where $D_{kpc}$ is the distance to M15 in kiloparsecs (Table~\ref{tab:m15_params}),
$\kappa_\nu$ is the dust absorption coefficient in cm$^2$ g$^{-1}$,
$B(\nu,T_d)$ is the Planck function in cgs units, and $T_d$ is the dust
temperature. $f_\nu$ at 70~\micron{} is 691.2
$\pm$ 61.1 mJy (Table~\ref{tab:obs_sf}), and $\kappa_\nu$ is taken from \citet{ossen94} to be 56
$\pm$ 11 cm$^2$ g$^{-1}$ at 70~\micron{}, assuming a standard MRN dust
distribution \citep{mrn} and an ISM-type composition consisting of
graphite and silicate grains. We derive a total dust mass
of $(9 \pm 2) \times 10^{-4}$~M$_\odot$, which agrees within a factor
of two compared to the value cited by \citet{evans03} of $(4.8 \pm
1.6) \times 10^{-4}$~M$_\odot$, and is approximately 2-4 times smaller
than the dust mass predicted by equation (1).  The discrepancy between
the {\it Spitzer} and {\it ISO} calculated dust masses may be largely
due to the different choice of $\kappa_\nu$, as the ICM flux densities
at 70~\micron{} agree within the errors stated
(Table~\ref{tab:obs_sf}).  $\kappa_\nu$ is the most uncertain
parameter in the dust mass estimate, as its value depends largly on
composition and density assumptions.  Therefore, we note that
$\kappa_\nu$ could be up to an order of magnitude larger than the
value we have invoked here.

The diffuse emission from ICM dust in M15 is located approximately
17\arcsec{} to the west of the cluster core. The paucity of diffuse
dust toward the cluster center, where the gravitational potential well
is steepest, is puzzling. One possible explanation for this asymmetry
is a collection of millisecond pulsars (PSRs) near the core of M15 
\citep{sun02}. Seven PSRs are located within 17\arcsec{} of the core 
(Fig.~\ref{fig:radio}), and the radiative environs associated with these 
objects could lead to destruction of dust grains by sputtering or other 
ablation processes and may also inhibit dust production in stellar
winds. The PSR nearest to the ICM dust distribution observed in the
MIPS image, PSR2129+1210F, is located on the northeast outer edge of
the 70~\micron{} emission \citep{taylor93}. If, on the other hand,
each mass-losing star has contributed 0.15~M$_\odot$ to the ICM on
average (assuming that each star will lose 0.3~M$_\odot$ over its
entire lifetime), then we see that an ICM dust mass of $1 \times
10^{-3}$~M$_\odot$ corresponds to mass lost from $\sim$10$^2$
stars. This suggests that the ICM is short-lived, as this many stars
will have passed through the AGB superwind phase, defined as the phase
in which the mass loss rate exceeds the nuclear burning mass
consumption rate (\citet{jacco99b} suggests $\dot{M} \sim 10^{-4} -
10^{-5}$~M$_\odot$~yr$^{-1}$ for low-mass stars), in only $\approx$
10$^{6}$~yr, which is much shorter than the cluster's relaxation timescale. The ICM dust
therefore cannot be expected to have relaxed and assumed the global shape of the
gravitational potential well. If this is the case, the offset of the dust
cloud from the center of the cluster would not come as a surprise.

Sources of the ICM dust include the post main sequence mass-losing stars 
identified in Fig.~\ref{fig:massloss}.  If we assume that the
dust-to-gas ratio scales in proportion to metallicity during the
superwind phase, as indicated by \citet{marshall04}, then at the metallicity of M15, a dust mass loss of
10$^{-10}$ M$_{\odot}$~yr$^{-1}$ is expected \citep{becker00}. With 
this mass loss rate, we again find that the dust has been accumulating for
approximately 1$\times$10$^{6}$ years,  significantly shorter than the time
between subsequent passages of the  galactic plane \citep{evans03}.
This short time scale suggests that dust  does not survive long in the
ICM.  Processes that could be responsible for  removing dust from the
ICM include ram pressure by the galactic halo gas,  radiation-driven
outflow or photo-destruction.

\section{CONCLUSIONS}
\label{sec:concl}

Analysis of our \textit{Spitzer} image data on the core of the globular 
cluster M15 show strong evidence for the presence of intercluster
medium (ICM) dust in the cluster core, with a mass of $(9 \pm 2)
\times 10^{-4}$~M$_\odot$ and with an equilibrium temperature of
$\approx 70$~K. This is the first secure, high signal to noise
detection of ICM dust in a globular cluster.  Also present surrounding
the core are populations of dusty AGB and post-AGB stars, along with
the planetary nebula K648. Using IRS spectral data, we have 
observed both the [\ion{S}{4}] and [\ion{Ne}{2}] fine structure lines
in K648 and have derived abundance estimates. 

The unique capabilities of {\it Spitzer} have enabled us to identify
both the interstellar dust and the dust producers in M15.  This is surprising
at such low metallicity ([Fe/H] = -2.4), and may have implications for dust production
in the early universe.  The mass of the ICM dust in M15 suggests that
it has been accummulating for $\sim 10^6$ years, which is a factor of
ten shorter than the time since the last galactic plane crossing. The
dust mass is also approximately 4 times smaller than the mass predicted by
\citet{evans03}. Both of these results imply
that such dust does not survive long compared to its production rate,
and is thus part of a stochastic process. 

\acknowledgments

We acknowledge helpful discussions with G. Schwarz and B. Moore regarding 
approaches to abundance modeling in PNe, and BDM for providing
source code
for multi-level atom calculations. M.~L.~B., C,~E.~W., L.~A.~H., E.~P., 
and R.~D.~G. are supported in part by NASA through \textit{Spitzer} 
contracts 1256406, 1215746, 1268006, and 1276760 issued by JPL/Caltech to the 
University of Minnesota, and by National Science Foundation grant 
AST02-05814. This publication makes use of data products from the Two
Micron All Sky Survey, which is a joint project of the University of
Massachusetts and the Infrared Processing and Analysis
Center/California Institute of Technology, funded by the National
Aeronautics and Space Administration and the National Science
Foundation.


\clearpage




\begin{figure}[h!]
\epsscale{1}

\figcaption{Three-color images of the core of M15 (NGC 7078) 
spanning 3.6~\micron{} to 
70~\micron{}. The field shown is 3' $\times$ 6.3'.  (a) Blue is 
3.6~\micron{}, green is 4.5~\micron{}, and red is 8~\micron{}.  The PN 
(K648) becomes visible near the upper left of the cluster core. 
(b) Blue is 5.8~\micron{}, green is 8~\micron{}, and red is 
24~\micron{}. The intra-cluster medium (ICM) becomes 
visible as two resolved high surface 
brightness patches (IR2 \& IR1b) off-center from the cluster
core. Three dusty sources, IR3a, IR3b, and IR4, also become prominent at 24~\micron{}.(c) 
Blue is 8~\micron{}, green is 24~\micron{}, and red is 70~\micron{}. 
The two ICM clouds are no longer resolved at 
70~\micron{} (IR1a). 
\label{fig:3color} }
\end{figure}

\clearpage

\begin{figure}[h!]
\epsscale{1}
\plotone{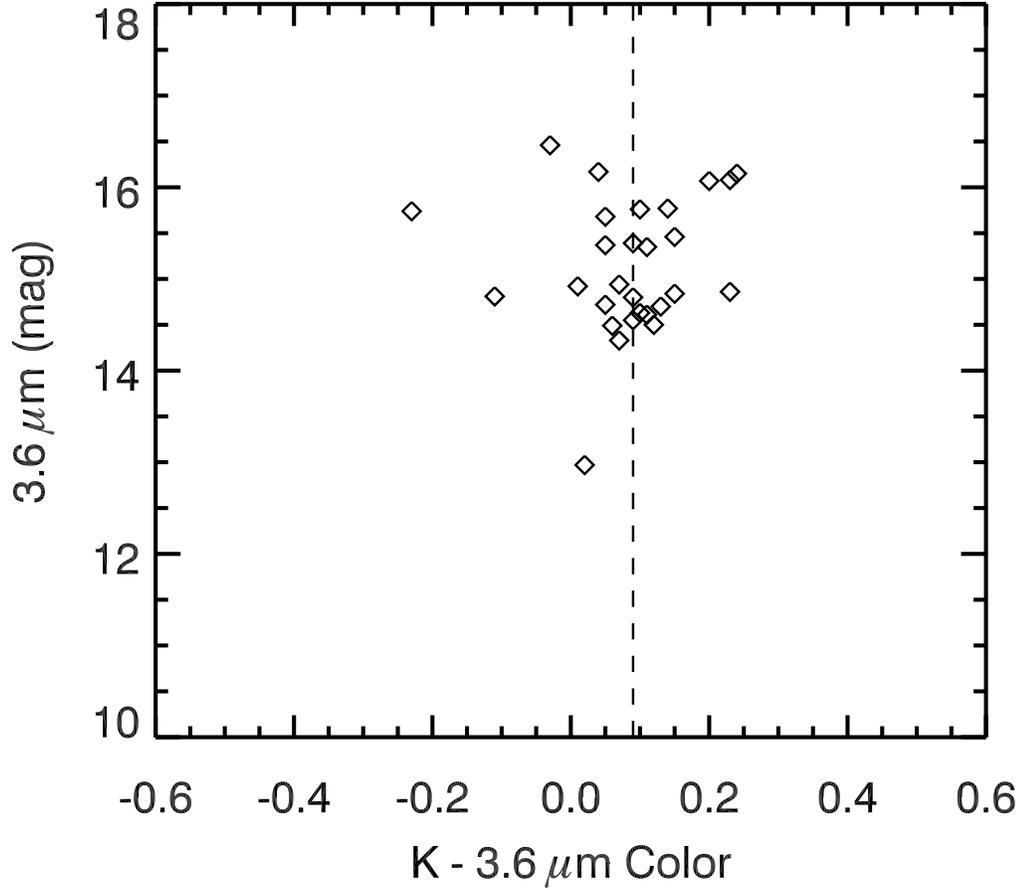}
\figcaption{Color magnitude diagram showing [K--3.6~\micron{}] colors.  K
magnitudes are from \citet{cohen05}.  The dashed line corresponds to
the median color, which is consistent with K giant stars.
\label{fig:cohen}}
\end{figure}

\clearpage

\begin{figure}[ht!]
\epsscale{.95}
\plotone{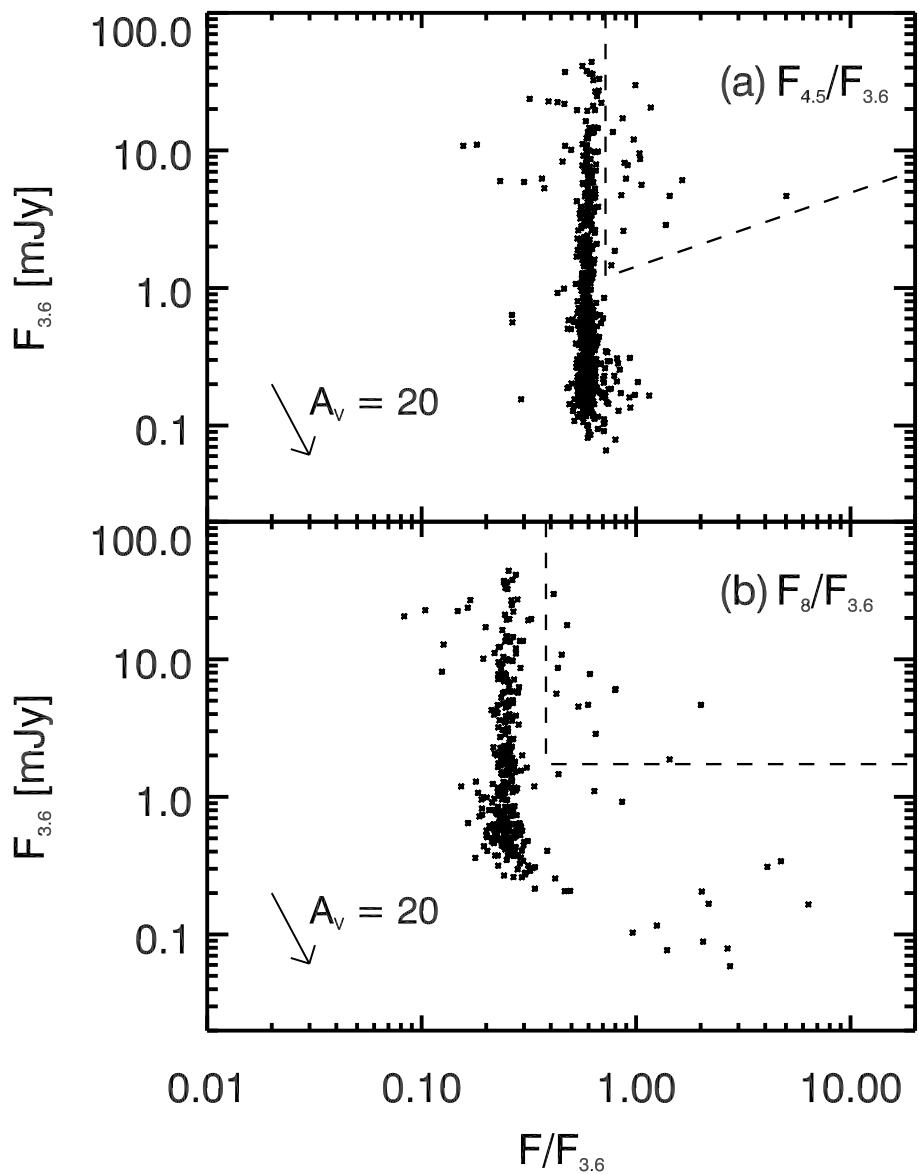}
\figcaption{IRAC color magnitude diagrams (CMDs) for M15. Fluxes are not 
color corrected or dereddened. (a) CMD for
4.5~\micron{}-3.6~\micron{} color vs 3.6~\micron{} flux. (b) CMD for
8~\micron{}-3.6~\micron{} color vs 3.6~\micron{} flux. The 
uncertainties in the photometry are given in Fig.~\ref{fig:err}.
The regions enclosed by the dashed lines (left of vertical, above
horizontal) are the mass loss regions.  The locations of these stars are
plotted in Fig.~\ref{fig:massloss}, and their fluxes are listed in
Table~\ref{tab:irsources}.
\label{fig:irac} }
\end{figure}

\clearpage

\begin{figure}[h!]
\epsscale{.95}
\plotone{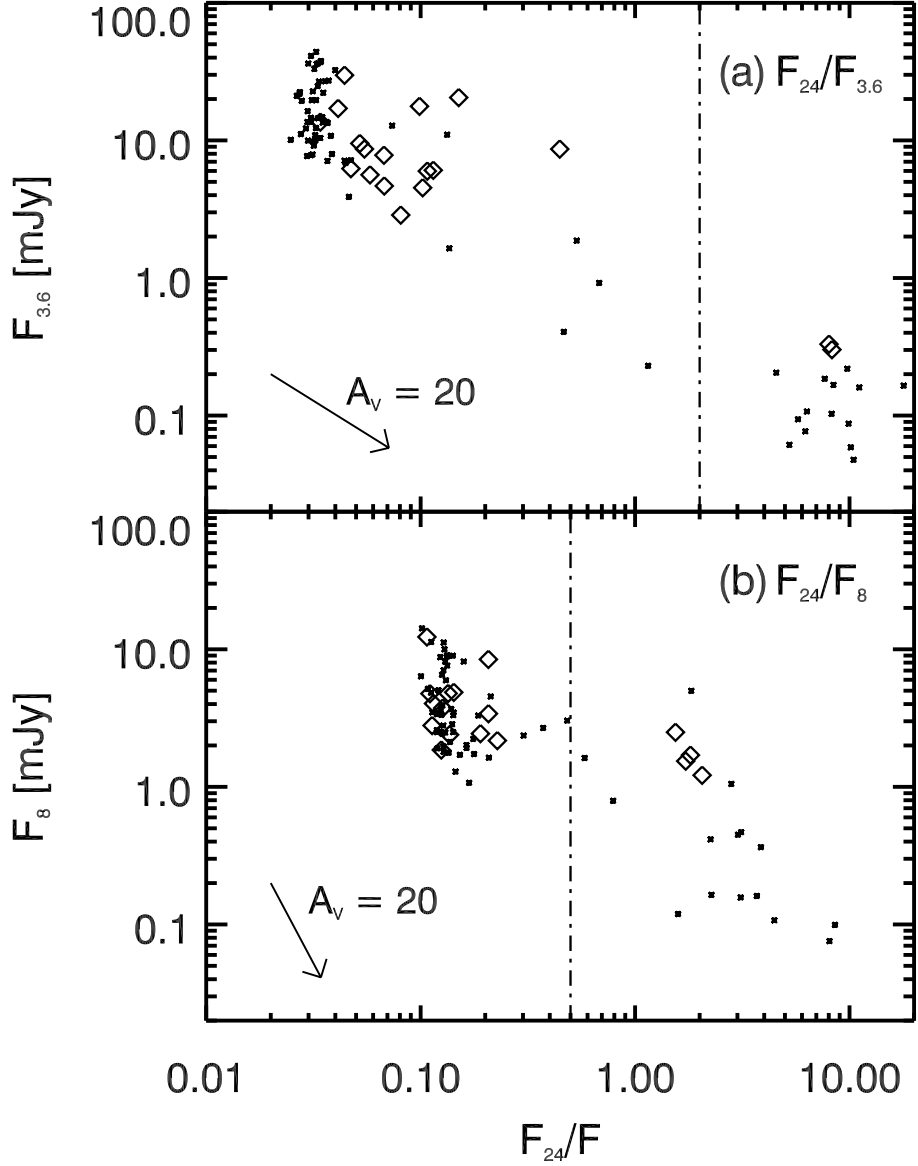}
\figcaption{MIPs color magnitude diagrams (CMDs) for M15. Fluxes are not
color corrected or dereddened.  (a) CMD for
24~\micron{}-3.6~\micron{} color vs 3.6~\micron{} flux. (b) CMD for
24~\micron{}-8~\micron{} color vs 8~\micron{} flux (see
Fig.~\ref{fig:err} for uncertainties in the photometry). The
diamonds mark the mass losing stars identified in Fig.~\ref{fig:irac}. 
The dash-dot lines mark the approximate division between AGB and 
post-AGB stars \citep{Groenewegen06}.  IR3a, IR3b, and IR4 each 
fall well within the post-AGB region in (b), while only IR3a and IR3b 
fall within the post-AGB region in (a). The IRAC colors of the other 
mass losing sources suggest that they are AGB stars.
\label{fig:mips} }
\end{figure}

\clearpage

\begin{figure}[h!]
\epsscale{0.6}

\figcaption{Photometry errors for the IRAC 3.6, 4.5, 8, and 
MIPS 24~\micron{} bands.
\label{fig:err} }
\end{figure}

\clearpage

\begin{figure}[h!]
\epsscale{1.0}
\plotone{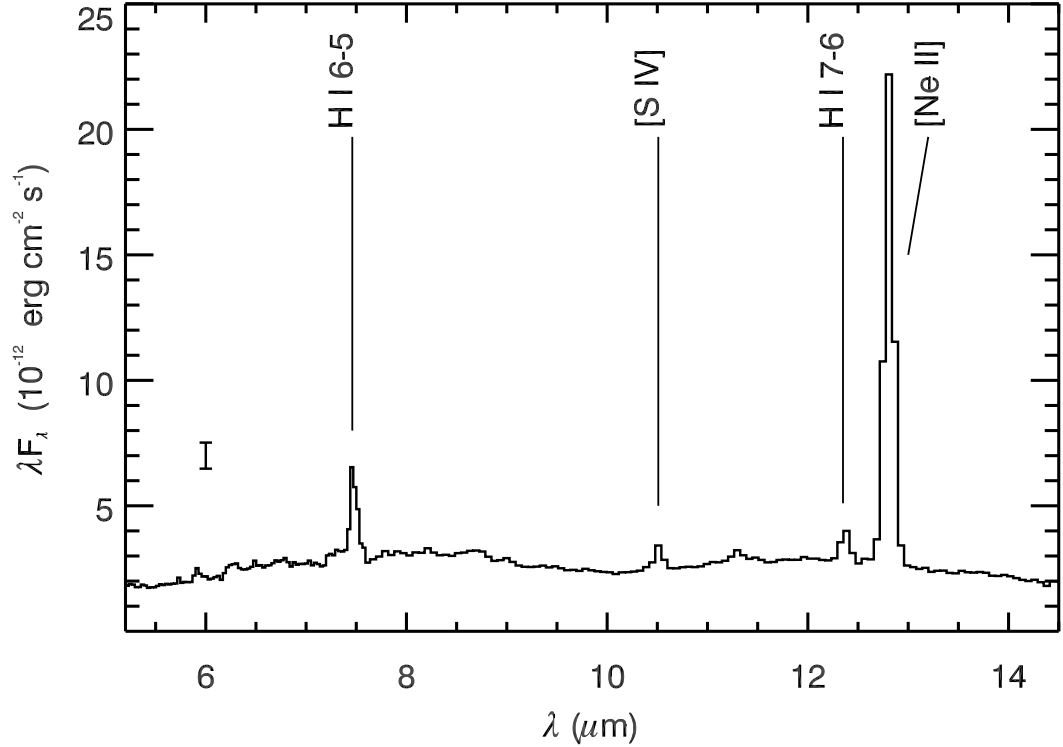}
\figcaption{\textit{Spitzer} IRS spectrum of the PN (K648) obtained with the short low (SL) module. The most prominent emission line in the spectrum is 
[\ion{Ne}{2}]$\lambda = 12.81~$\micron \, in addition to weak H 
recombination lines and the [\ion{S}{4}] line. The representative error bar 
({\it left inset}) shows the average photometric uncertainty in the spectra.
No strong dust continuum is evident in the IRS spectrum.
\label{fig:pn} }
\end{figure}

\clearpage

\begin{figure}[h!]

\epsscale{0.8}

\figcaption{Three color image of 8~\micron{} (blue), 24~\micron{}
(green), and 70~\micron{} (red)
images. The 24~\micron{} and 8~\micron{} images were rebinned to the
70~\micron{} image pixel size of 4.925 arcseconds then convolved
with the 70~\micron{} \textit{Spitzer} \ point response function (PRF). This 
image shows that the 70~\micron{} emission is not unresolved stellar emission.
\label{fig:convolved} }
\end{figure}

\clearpage

\begin{figure}[h!]
\epsscale{1.1}
\figcaption{Locations of Mass-losing stars in M15. (a) the
3.6~\micron{} mosaic and (b) the 24~\micron{} mosaic. All of
the stars marked are located redward of the RGB.
\label{fig:massloss} }
\end{figure}

\clearpage

\begin{figure}[h!]
\epsscale{1}
\plotone{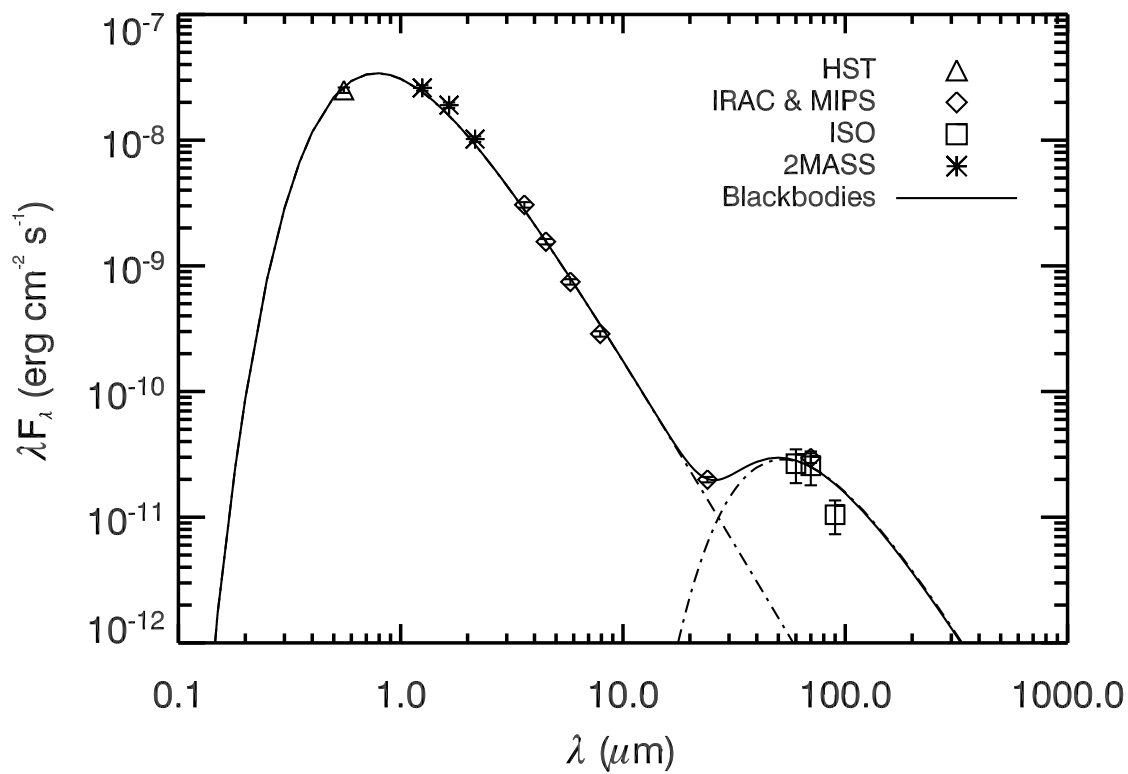}
\figcaption{Spectral energy distribution (SED) of the stellar and
dust components near the core of M15.  Fluxes (see Table~\ref{tab:obs_sf}) 
were obtained by using a square aperture centered on the dust emission
at \textit{ISO} \  wavelengths. The best 
two-blackbody fit ($\chi^2$ = 5.26) yields 4698
$\pm$ 58~K and 70 $\pm$ 2~K for the stellar and dust components,
respectively. The dash-dot lines are the individual blackbodies, while
the solid line is the sum of the two blackbodies.
\label{fig:sed} }
\end{figure}

\clearpage

\begin{figure}[h!]
\epsscale{0.8}
\plotone{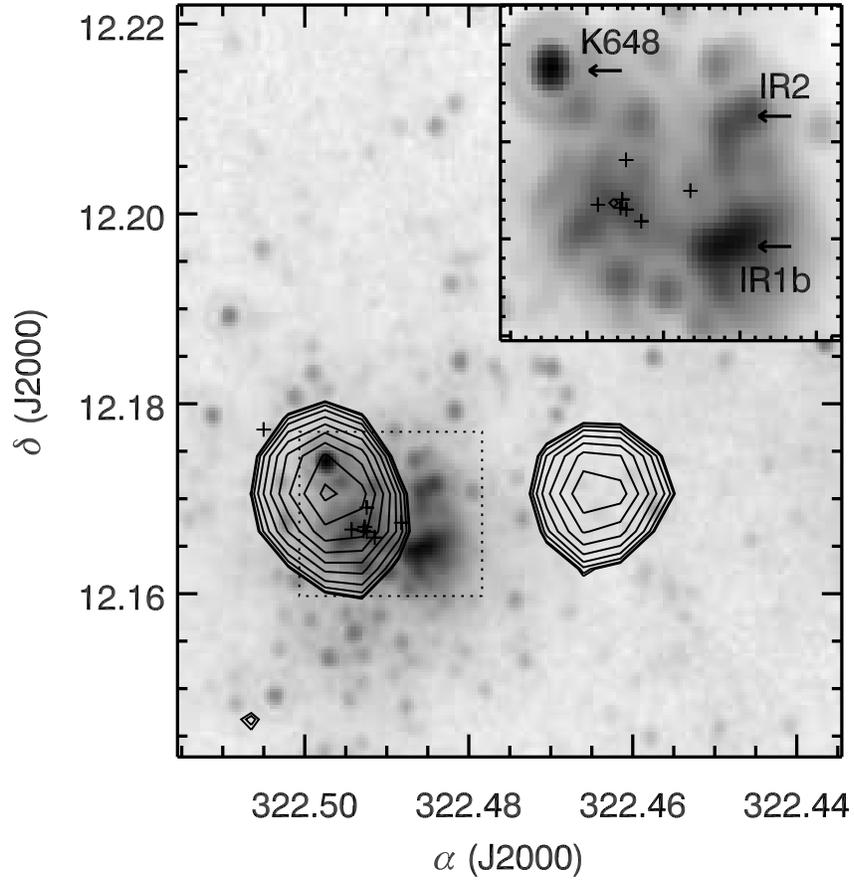}
\figcaption{MIPS 24~\micron{} mosaic overlayed with radio contours
  (1.4~GHz) \citep{condon98}. There are
  ten contour
  levels ranging from pixel values of 1.56 to 5.0 mJy/Beam on a square
  root scale.  Pulsar locations are marked with an ``$+$'' symbol, and
  the center of the cluster is marked by a diamond.  The inset (dotted
  lines) has dimensions of approximately 1.3\arcmin{} $\times$
  1.0\arcmin{}.  
\label{fig:radio} }
\end{figure}

\clearpage

%



\begin{deluxetable}{ccccccccc}
\tablewidth{0pc}
\tablecaption{M15 properties\label{tab:m15_params}\tablenotemark{a}}
\tablehead{\colhead{[Fe/H]} & \colhead{Core radius} &
  \colhead{Z-height} & \colhead{$\tau_c$} & \colhead{M$_{cluster}$} &
  \colhead{V$_{esc}$} & \colhead{L} & \colhead{E(B-V)} & \colhead{D}\\
  & \colhead{('')} & \colhead{(kpc)} & \colhead{(years)} &
    \colhead{(M$_\odot$)} & \colhead{(km s$^{-1}$)} &
    \colhead{(L$_\odot$)} & & \colhead{(kpc)}}
\startdata
-2.4 &4.1&4.79&4x10$^7$&2x10$^6$&40.9&5.8x10$^5$&0.11&9.98$\pm$0.47\\
\enddata
\tablenotetext{a}{ \ Values reproduced from \citet{evans03}, except E(B-V)
  from \citet{schlegel98} and distance from \citet{mcnamara04}.}

\end{deluxetable}


\begin{deluxetable}{crccc}
\tablewidth{0pc}
\tablecaption{Observational Summary \label{tab:obsum}}
\tablehead{\colhead{Instrument} & \colhead{AORkey} & \colhead{Exposure
    Time\tablenotemark{a}} & \colhead{Coverage} & \colhead{pixel
    size}\\
&&\colhead{(s)}&&\colhead{(''/pixel)}}
\startdata

IRAC & 12030208 & 10.4 & 9.5' x 9.5' & 1.22 \\
MIPS 24~\micron{}& 12030464 & 9.96 & 11' x 12' & 1.245 \\
MIPS 70~\micron{}& 12030464 & 10.49 & 8.5' x 14.5' & 4.925 \\
IRS & 15733760 & 60 (SL) \\
\enddata
\tablenotetext{a}{ \ Exposure time of individual BCD images. For IRAC,
  only medium exposure HDR frames are listed.}
\end{deluxetable}

%

\begin{deluxetable}{ccccrcrcrcrcr}
\rotate
\tablewidth{0pc}
\tablecaption{Flux Densities\tablenotemark{a} \ of Dusty IR Sources\label{tab:irsources}}
\tablehead{\colhead{ID\tablenotemark{b}}&&\colhead{r}&&\colhead{3.6~\micron{}}
    &&\colhead{4.5~\micron{}}&&\colhead{5.8~\micron{}}&&\colhead{8~\micron{}}&&\colhead{24~\micron{}}\\
&&\colhead{arcsec}&&\multicolumn{9}{c}{mJy}}
\startdata

2MASS J21295231+1210515  &&0.212 &&13.99$\pm$0.70&&10.87$\pm$0.54&& 7.02$\pm$0.35&& 4.22$\pm$0.21&&0.44$\pm$0.02\\

&&&&&&&&&&&\\

2MASS J21295264+1210440  &&   0.201 && 6.24$\pm$0.31&&10.21$\pm$0.51&& 8.64$\pm$0.43&& 5.09$\pm$0.25&&0.66$\pm$0.02\\

&&&&&&&&&&&&\\

SSTU J212953.55+120910.7 & &  0.982 && 8.91$\pm$0.45&& 9.28$\pm$0.46&& 6.80$\pm$0.34&& 3.91$\pm$0.20&&0.45$\pm$0.02\\

&&&&&&&&&&&&\\

2MASS J21295383+1209338 & &  0.200   && 6.15$\pm$0.31&& 1.43$\pm$0.07&& 5.22$\pm$0.26&& 5.00$\pm$0.25&&0.61$\pm$0.02\\

&&&&&&&&&&&&\\

2MASS J21295473+1208592  &&   0.032  && 4.80$\pm$0.24&& 6.87$\pm$0.34&& 5.23$\pm$0.26&& 2.92$\pm$0.15&&0.30$\pm$0.02\\

&&&&&&&&&&&&\\

2MASS J21295618+1210179  &&   0.030 &&21.08$\pm$1.05&&24.61$\pm$1.23&&22.05$\pm$1.10&& $>$1.78$\pm$0.09&&2.93$\pm$0.02\\

&&&&&&&&&&&&\\

2MASS J21295678+1210269  &&   0.271 &&30.65$\pm$1.53&&30.35$\pm$1.52&& $>$6.19$\pm$0.31&&12.86$\pm$0.64&&1.25$\pm$0.02\\

&&&&&&&&&&&&\\

2MASS J21295703+1209376  & &  0.242 && 4.79$\pm$0.24&&23.99$\pm$1.20&&17.84$\pm$0.89&& 9.78$\pm$0.49&&-\\

&&&&&&&&&&&&\\

2MASS J21295712+1210043  &&   0.190   &&11.11$\pm$0.56&& 1.73$\pm$0.09&& 4.79$\pm$0.24&& 5.08$\pm$0.25&&-\\

&&&&&&&&&&&&\\

2MASS J21295716+1209175  &&   0.250   &&17.59$\pm$0.88&&15.18$\pm$0.76&&12.41$\pm$0.62&& 3.56$\pm$0.18&&0.67$\pm$0.02\\

&&&&&&&&&&&&\\

2MASS J21295758+1209552  &&   0.301   &&12.34$\pm$0.62&&12.00$\pm$0.60&& $>$0.27$\pm$0.01&& 3.45$\pm$0.17&&-\\

&&&&&&&&&&&&\\

2MASS J21295756+1210276   &&   0.022  && 6.06$\pm$0.30&& - && $>$0.62$\pm$0.03&& 2.55$\pm$0.13&&0.44$\pm$0.02\\

&&&&&&&&&&&&\\

2MASS J21295815+1209466  &&   0.493 &&18.20$\pm$0.91&&28.30$\pm$1.41&&21.12$\pm$1.06&& 8.83$\pm$0.44&&1.66$\pm$0.02\\

&&&&&&&&&&&&\\

2MASS J21295828+1209280 &&   0.202 &&4.85$\pm$0.24&& 4.13$\pm$0.21&& 0.91$\pm$0.05&& 0.33$\pm$0.02&&-\\

&&&&&&&&&&&&\\

2MASS J21295832+1209128  &&   0.181   && 8.03$\pm$0.40&& 7.32$\pm$0.37&& 5.40$\pm$0.27&& 4.98$\pm$0.25&&0.50$\pm$0.02\\

&&&&&&&&&&&&\\

2MASS J21295881+1209285  &&   0.301 && 9.78$\pm$0.49&&10.12$\pm$0.51&& 5.39$\pm$0.27&& 2.27$\pm$0.11&&0.47$\pm$0.02\\

&&&&&&&&&&&&\\

2MASS J21295937+1210029  &&   0.420 && 8.35$\pm$0.42&& 7.35$\pm$0.37&&8.61$\pm$0.43&& 1.06$\pm$0.05&&-\\

&&&&&&&&&&&&\\
 
SSTU J212959.69+120739.0\tablenotemark{c}&&   4.722   &&0.31$\pm$0.04&&0.29$\pm$0.02&&0.26$\pm$0.02&&1.47$\pm$0.07&&2.37$\pm$0.02\\

&&&&&&&&&&&&\\

2MASS J21295981+1211107  &&   0.231  && 6.41$\pm$0.32&& 5.73$\pm$0.29&& 4.14$\pm$0.21&&-&&0.28$\pm$0.02\\

&&&&&&&&&&&&\\

2MASS J21295996+1207282\tablenotemark{d} & &   0.100    &&0.34$\pm$0.03&&0.25$\pm$0.02&&0.24$\pm$0.02&&2.05$\pm$0.10&&2.52$\pm$0.03\\

&&&&&&&&&&&&\\

2MASS J21300062+1209284  &&   0.140   && 2.95$\pm$0.15&& 4.05$\pm$0.20&& $>$0.12$\pm$0.01&& 1.94$\pm$0.10&&0.22$\pm$0.02\\

&&&&&&&&&&&&\\

2MASS J21300097+1210375  &&   0.054  && 5.78$\pm$0.29&& 6.11$\pm$0.31&& 4.41$\pm$0.22&& 2.50$\pm$0.12&&0.31$\pm$0.02\\

&&&&&&&&&&&&\\

2MASS J21300277+1206557\tablenotemark{e} & &   0.010  &&9.34$\pm$0.47&&6.18$\pm$0.31&&4.04$\pm$0.20&&3.29$\pm$0.16&&3.66$\pm$0.02\\

&&&&&&&&&&&&\\

\enddata

\tablenotetext{a}{Flux densities are color corrected and dereddened
  using E(B-V) = 0.11 and A$_\lambda$ values from \citet{indebetouw05}.}
\tablenotetext{b}{Standard 2MASS designations are listed for all stars
  that have a 2MASS counterpart located within r = 0.5\arcsec~from the
  \textit{Spitzer} coordinates.  Otherwise, stars are listed with
  standard \textit{Spitzer} designations where U stands for ``unidentified''.}
\tablenotetext{c}{Designated as IR3a in text. 3.6, 4.5, and
  5.8~\micron fluxes were determined with aperture photometry in IDL,
  as this source was not detected by APEX at these wavelengths.}
\tablenotetext{d}{Designated as IR3b in text. 3.6, 4.5, and
  5.8~\micron fluxes were determined with aperture photometry in IDL,
  as this source was not detected by APEX at these wavelengths.}
\tablenotetext{e}{Designated as IR4 in text.}

\end{deluxetable}

%

\begin{deluxetable}{rrrc}
\tablewidth{0pc}
\tablecaption{ICM Dust (IR1a) Flux Densities \label{tab:obs_sf}}
\tablehead{\colhead{$\lambda$}& & \colhead{F$_{\lambda}$\tablenotemark{a}} &
    \colhead{Instrument(s)/Mission(s)} \\
\colhead{\micron} && \colhead{mJy} &}
\startdata
0.56&&4667.$\pm$233.&HST WFPC2\tablenotemark{b} \\
1.25&&10857.$\pm$186.&2MASS\\
1.65&&10450.$\pm$204&2MASS\\
2.16&&7342.$\pm$134.&2MASS\\
3.6&&\textbf{3669.$\pm$1.}&\textbf{IRAC}\\
4.5&&\textbf{2336.$\pm$1.}&\textbf{IRAC}\\
5.8&&\textbf{1442.$\pm$1.}&\textbf{IRAC}\\
8.0&&\textbf{757.7$\pm$0.5}&\textbf{IRAC}\\
24&&\textbf{159.4$\pm$0.1}&\textbf{MIPS}\\
60&&515.$\pm$155.&ISO\tablenotemark{c}\\
70&&\textbf{691.2$\pm$61.1}&\textbf{MIPS}\\
70&&578.$\pm$173.&ISO\tablenotemark{c}\\
90&&303.$\pm$91.& ISO\tablenotemark{c}\\

\enddata
\tablenotetext{a}{ \ Aperture size is 130.5 $\times$ 130.5 arcseconds,
  corresponding to the ISO aperture from \citet{evans03}. All fluxes are dereddened 
with E(B-V)=0.11. A$_\lambda$ values come from \citet{Rieke85} and \citet{indebetouw05}.}
\tablenotetext{b}{ \ From image U5B80101R from the public HST
  archives.}
\tablenotetext{c}{ \ From \citet{evans03}.}

\end{deluxetable}


\begin{deluxetable}{rrlrr}
\tablewidth{0pc}
\tablecaption{Spitzer Infrared Line Fluxes of K648 \label{tab:k648lf}}
\tablehead{\colhead{$\lambda$} && \colhead{ } && \colhead{Flux\tablenotemark{a}} \\
           \colhead{\micron{}} && \colhead{Ion/Line}&& \colhead{ergs cm$^{-2}$ s$^{-1}$}}
\startdata

12.81&&[\ion{Ne}{2}]&& $(1.74 \pm 0.08) \times 10^{-13}$ \\

12.35&&\ion{H}{1} $7 - 6$&& $(9.75 \pm 3.08) \times 10^{-15}$\\

10.51&&[\ion{S}{4}]&& $(8.24 \pm 3.56) \times 10^{-15}$\\

7.46&&\ion{H}{1} $6 - 5$&& $(4.38 \pm 0.68) \times 10^{-14}$\\
\enddata
\tablenotetext{a}{ \ Integrated line fluxes assuming a fixed IRS 
instrumental resolution of 0.0404~\micron , and a Gaussian line profile.} 
\end{deluxetable}

\begin{deluxetable}{llcrr}
\tablewidth{0pc}
\tablecaption{Sulfur and Neon Abundances for K648 \label{tab:k648abund}}
\tablehead{\colhead{Quanity}&&&& \colhead{Value}}
\startdata

$T_{e}$(K) &&&& 12,500 \\

$n_{e}$(cm$^{-3}$) &&&& 1,700 \\

$I$([\ion{S}{2}]6717,6731)/$I$(H$\beta$)\tablenotemark{a}&&&& $(1.69 \pm 1.12) \times 10^{-3}$ \\

S$^{+}$/H$^{+}$&&&&$(6.30 \pm 3.00) \times 10^{-9}$\\

$I$([\ion{S}{3}]9532)/$I$(H$\beta$)\tablenotemark{b}&&&& $(7.59 \pm 3.37) \times 10^{-3}$\\

S$^{2+}$/H$^{+}$ &&&& $(2.55 \pm 1.13) \times 10^{-8}$ \\

$I$([\ion{S}{4}]10.51~\micron{}/$I$(H$\beta$)\tablenotemark{c}&&&& $(5.42 \pm 2.54) \times 10^{-3}$\\

S$^{3+}$/H$^{+}$&&&&$(1.10 \pm 0.52) \times 10^{-8}$\\

Total S/H &&&&$ (4.28 \pm 1.28) \times 10^{-8}$\\

$I$([\ion{Ne}{3}]3967)/$I$(H$\beta$)\tablenotemark{d}&&&& $(2.7 \pm 0.9) 
\times 10^{-3}$\\ 

Ne$^{2+}$/H$^{+}$ &&&&$(3.69 \pm 1.23) \times 10^{-6}$\\

$I$([\ion{Ne}{3}]3969)/$I$(H$\beta$)\tablenotemark{d}&&&& $(1.19 \pm 0.30) \times 10^{-1}$\\

Ne$^{2+}$/H$^{+}$ &&&&$(4.89 \pm 1.23) \times 10^{-6}$\\

$I$([\ion{Ne}{2}]12.81~\micron{}/$I$(H$\beta$)\tablenotemark{c}&&&& $(1.15 \pm 0.18) \times 10^{-1}$\\

Ne$^{+}$/H$^{+}$&&&& $(1.53 \pm 0.21) \times 10^{-5}$\\

Total Ne/H &&&&$(2.39 \pm 0.27) \times 10^{-5}$\\

O/H\tablenotemark{e}&&&& $(5.0 \pm 1.3) \times 10^{-5}$\\

[S/O]\tablenotemark{f}&&&& --2.64\\

[Ne/O] \tablenotemark{f}&&&& +0.54\\

\enddata
\tablenotetext{a}{\ From \citet{barker83}.}
\tablenotetext{b}{\ From \citet{barker83}. Note lines are blended.}
\tablenotetext{c}{\ H$\beta$ flux derived from \textit{Spitzer} observations of \ion{H}{1} recombination lines (see \S\ref{sec:hbeta}).}
\tablenotetext{d}{\ From \citet{adams84}. Measured values corrected 
for extinction (see \S\ref{sec:neona}).}
\tablenotetext{e}{\ From \citet{pena93}.}
\tablenotetext{f}{$[X] = log_{10}(X_{Object}) - log_{10}(X_{\odot})$.}
\end{deluxetable}

\end{document}